# FPGA Based Accelerator for Neural Networks Computation with Flexible Pipelining

Qingyang YI[1]   Heming SUN[23]   and   Masahiro FUJITA[4]

[1]Graduate School of Engineering, The University of Tokyo   7-3-1 Hongo, Bunkyo-ku, Tokyo, 113-8654 Japan
[2]Waseda Research Institute for Science and Engineering, Waseda University 3-4-1 Ookubo, Shinjuku, Tokyo, 169-8555
[3]JST, PRESTO   4-1-8 Honcho, Kawaguchi, Saitama, 332-0012
[4]System Design Research Center, University of Tokyo   7-3-1 Hongo, Bunkyo-ku, Tokyo, 113-8654 Japan

E-mail:   [1]yiqi@cad.t.u-tokyo.ac.jp,   [2]hemingsun@aoni.waseda.jp,   [4]fujita@ee.t.u-tokyo.ac.jp

**Abstract**   FPGA is appropriate for fix-point neural networks computing due to high power efficiency and configurability. However, its design must be intensively refined to achieve high performance using limited hardware resources. We present an FPGA-based neural networks accelerator and its optimization framework, which can achieve optimal efficiency for various CNN models and FPGA resources. Targeting high throughput, we adopt layer-wise pipeline architecture for higher DSP utilization. To get the optimal performance, a flexible algorithm to allocate balanced hardware resources to each layer is also proposed, supported by activation buffer design. Through our well-balanced implementation of four CNN models on ZC706, the DSP utilization and efficiency are over 90%. For VGG16 on ZC706, the proposed accelerator achieves the performance of 2.58x, 1.53x and 1.35x better than the referenced non-pipeline architecture [1], pipeline architecture [2] and [3], respectively.

## 1. INTRODUCTION

Inference of Convolution Neural Networks (CNN) is a complicated computation task consisting of up to several billion of multiplication and accumulation (MAC) operations. Compared with general information processing, dedicated design of hardware, which is called neural networks (NN) accelerator, is able to achieve high performance on NN applications. There are critical requirements when considering NN accelerator design, such as high throughput, low latency, power efficiency and area constraint. However, there are a variety of NN models of which features are quite different, such as kernel size, number of layers and channels. Because the complexity and data amount are also different, it is difficult to develop a single general architecture, which keeps high efficiency for most NN models. Therefore, NN accelerator should have reconfigurability to adjust to specific NN model.

Field Programmable Gate Array (FPGA) is a category of integrated circuit with plenty of logic units, memory units, and programmable routing resources. FPGA is intrinsically reconfigurable so that it can accommodate various NN models. Besides, FPGA can achieve good efficiency especially for fixed-point NN computation. General CPU is able to achieve flexible control but the limited computation units and poor power efficiency limit efficiency of NN applications. GPU possesses lots of computation units but may not be able to achieve energy efficient computations. [4] compares CPU, GPU, FPGA through evaluations on the implementation of RNN models, and illustrates that FPGA achieved much higher hardware utilization and power efficiency than CPU/GPU across different hidden layer sizes, and realized higher performance. Compared with ASIC design, FPGA has faster design and verification cycle and omits high fabrication cost. Therefore, we choose FPGA as target platform for NN accelerator implementation. It is worthy to note that hardware resources of FPGA (LUTs, FFs, BRAMs, DSPs and DDR bandwidth) are limited for specific FPGA. Therefore, when working on FPGA design, how to make high utilization of limited resources is the key to achieve high performance. Besides, the design should be logic optimized for higher clock speed.

There are two categories of NN accelerator architectures: recurrent and pipeline according to [5]. The recurrent architecture consists of a layer-specific processing element array, which processes the layers one-by-one, while the pipeline architecture processes the multiple layers on-chip simultaneously. The pipeline architecture is preferable over recurrent architecture regarding FPGA implementation. First of all, the pipeline architecture has potential of higher hardware utilization because hardware resources allocation can be customized according to characteristics of each layer. Secondly, we can balance hardware resources assignment among layers to reduce idle cycles of DSPs, for higher hardware efficiency. Thirdly, the pipeline architecture is able to save the power by avoiding activation transfer between on-chip and off-chip when processing different layers.

In this paper, we present an FPGA based pipeline architecture for NN inference. The convolution layer is implemented in parameterized style so that parallelism can be easily adjusted. We design a flexible activation line buffer which transfers activation data between two consecutive convolution layers with significant different parallelism. Based on this architecture, we propose the hardware resource allocation algorithm and its framework, which can customize flexible pipeline accelerator for given NN model and FPGA board.

This paper is organized as follows. Section 2 gives the basics of NN accelerator design and compare recurrent and pipeline architecture by referring to related works. Section 3 and 4 introduce our proposed architecture and the hardware resources allocation algorithm. The setup of the experiments and the results are presented in Section 5. Section 6 concludes this work.

## 2. RELATED WORKS

### 2.1. NN Accelerator

CNN is named from convolution layers, which has good ability for feature extraction so that CNN is widely used in image processing, video compression and data analysis. A convolutional layer performs convolution computation on input feature map($I$) panel with weight($W$) and generate output feature map($O$), as follows:

$$O_{M \times H \times W} = f(W_{M \times C \times R \times S} \circledast I_{C \times (H+R-1) \times (W+S-1)} + B_M) \quad (1)$$

The number of CNN layers can vary from a few layers to tens of layers, and numbers of channels are usually several hundreds, which makes CNN a computation-dense task. For example, the complexity of VGG16, a typical image processing CNN, is 30.94 GOP. Besides, the data amount of weight data and internal results are quite large.

In order to run CNN inference efficiently, which possesses large computation workload and data amount, a sophisticated architecture for neural network model computation is necessary, which is called Neural Network Accelerator. The core part of NN Accelerator is processing element array which mainly contains multipliers and adders to deal with MAC operations. Some buffers are implemented by on-chip memories to store data such as input activations, weights and partial sums. However, the amount of these data is too large to store on chips entirely, so most data is stored in off-chip memories such as DDR, and buffers load small portions of data incrementally for individual current computations. The off-chip memory access is much more power-consuming than on-chip memory access and consumes off-chip memory access bandwidth. Good news is that convolution layer has great potential for data reuse. In convolution layers, each input activation or weight is associated with multiple computations and can be reused after being loaded on chip so as to reduce off-chip access. Many researches concentrate on the design of dataflow, which decides how the computations are organized to reach higher data reuse. Eyeriss [6] architecture adopts row-stationary dataflow and designs with networks-on-chip, so that both activation data and weight data are highly reused.

To design an architecture to compute most NN models efficiently is a difficult challenge due to the variety of NN models. Firstly, the numbers of layers and channels of each layer are quite different so that they have different computation workload. Secondly, computation to communication (CTC) ratio varies among NN models or layers of a model. One architecture can achieve optimal performance only for several models with similar CTC ratio. Thirdly, the kernel size and activations size may be different, which affects the design of dataflow. NN accelerators should hold flexibility for various models.

Besides, there are many other requirements when considering a NN accelerator design, due to the deployment of NN applications on various environments, such as throughput, latency, power efficiency and resources constraints.

### 2.2. Comparison of Architecture – Recurrent and Pipeline

There are two categories of NN accelerator architectures: recurrent and pipeline. Both of them are able to be used as FPGA-based accelerator architecture.

Recurrent architecture possesses a layer-specific processing element array, which is reused for processing all CNN layers consecutively, one layer at a time. Internal activation data is buffered by off-chip memory when the size of feature map is too large to store on-chip. A dedicated mapping is required to distribute and assign computation workload to each processing element on NN accelerator. Generally, it is accomplished by a compiler or by hand, and corresponding a sort of instruction set architecture (ISA) design is necessary. A configurable NoC is usually required as connections of processing elements to make the fixed array adaptive and reusable for different CNN layers. DSIP [7] proposes a processing element design called DALU for recurrent convolution operations. An instruction set [8][9] is designed to support convolution task mapping. All convolutions are still processed by a fixed-shaped processing element array, so that the performance is not always best depending on NN model. For example, an array of 9 PEs is not able to handle 5x5 kernel and can handle double 2x2 kernel but 1 PE is idle.

Pipeline architecture processes multiple layers on-chip at the same time, and activation data is passed by on-chip buffers between two consecutive layers. Hardware resources are assigned to each layer according to the workload and hyperparameters of related convolution layers so that the performance can be optimal for any specific layer. [2] proposes a fusion pipeline architecture, combining convolution layer implemented by conventional and Winograd algorithm together. Allocation algorithm targeting on low latency achieves relatively low hardware efficiency. DNNBuilder [3] is a representative pipeline architecture. All layers are implemented on FPGA and connected by activation line buffer as input activation cache. Activations are computed and passed between layers in a column-based scheme. Processing element array of each layer has parameterized parallelism so that computation cycles for one column can be adjusted. Based on that, resource allocation is performed in order to balance speed of all layers so that high hardware efficiency can be achieved. However, due to defect of activation buffers design, resource allocation algorithm has such constraints: input channel parallelism of current layer must be equal to output channel parallelism of the previous layer and channel parallelism must be power of 2 for saving BRAM of activation buffers. Therefore, hardware utilization is not so satisfactory.

To conclude, pipeline architecture can achieve higher hardware efficiency than recurrent architecture. If we can allocate hardware to each layer freely, supported by more flexible hardware design, pipeline architecture is also able to make full use of more hardware resources. The requirement of pipeline architecture is that circuit must change with NN models. Although it may not be appropriate for ASIC, the reconfigurability of FPGA can accommodate such requirement.

## 3. ARCHITECTURE
### 3.1. Dataflow
CNN model computation can be represented by a series of nested for-loops, where each for-loop processes one of multiple dimensions: layers $L$, input channels $C$, output channels $M$, feature maps' height $H$, feature maps' width $W$ and kernel size $R \times S$. Neural computation architectures usually optimize the nested for-loop by flattening, unrolling, pipelining and so on, so that hardware-friendly optimal dataflows can be established.

```
for l in range(L): #layers
  for h in range(H/K): #frame height
    for mm in range(M/M'): #output channel group
      for cc in range(C/C'): #input channel group
        for k in range(K): #row parallelism
          for m in range(M'): #output channel parallelism
            for c in range(C'): #input channel parallelism
              for w in range(W): #frame width
                for r in range(R): #kernel rows
                  for s in range(S): #kernel width
                    oact[l][h*K+k][w][mm*M'+m]
                    += weight[l][r][s][mm*M'+m][cc*C'+c]
                    * iact[l-1][h*K+k+r][w+s][cc*C'+c]
```
Fig.1 Dataflow of the proposed architecture

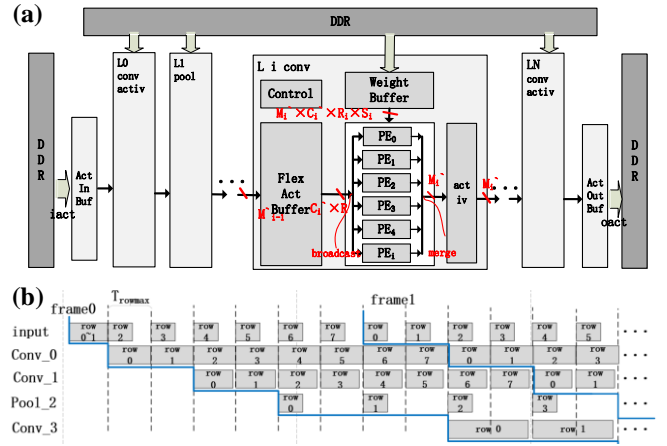

Fig.2 The proposed architecture design

Our dataflow design follows these principles: layer-wise pipeline, kernel/channel-wise PE parallelism, and higher weight reuse for lower off-chip memory access bandwidth. Fig.1 shows pseudo-code of our optimized CNN model computation. Notice that, the activation rows, output channels and input channels are divided into $K$-row groups, $M'$-channel groups and $C'$-channel groups, which are called row parallelism, input channel parallelism and output channel parallelism respectively. Weight data is reused by the computation of $K$ rows of activations so that off-chip memory access bandwidth can be smaller. $M'$ and $C'$ are adjustable parameters for each convolution layer engine and decide the number of processing elements. Once a convolution layer engine has $M' \times C' \times R \times S$ weight data and $(R + (K-1) \times S) \times W \times C'$ activation data in the buffers, computation is in progress. Then computation of new channels starts and the computation of new activation row is scheduled after all channels of the current row has been accomplished. Regarding of transfers between layers, the atomic element of activation data is of $K \times W \times C$ size, which flows through the pipeline consisting of all layers. Each convolution layer engine performs $M' \times C' \times R \times S$ MAC operations in one cycle.

## 3.2. Pipeline Top

The proposed architecture is shown in Fig.2. Input activation and weight data are stored in the off-chip DDR initially in binary format. To address the latency of DDR access, actIn buffer and weight buffer are designed to keep these data and unpack from binary to pixels. Correspondingly, actOut buffer packs output activation data and sends back to DDR.

All layers of a NN model are implemented on chip simultaneously and connected in sequence. Major layers, including convolution layers, pooling layers and full-connected layers, are implemented as individual pipeline stages because these layers dominate the majority of the computation and input data must be buffered for reuse.

The number of cycles for convolution layer engine i to compute $K$ activation rows is calculated as follows:

$$T_{rowi} = K_i \times W_i \times \frac{C_i}{C'_i} \times \frac{M_i}{M'_i} \qquad (2)$$

As Fig.1(b) shows, if $T_{row}$ of all layers are different, faster layers may be idling until slower layers finish computations. The number of cycles of pipeline stage is determined by the slowest layer, termed as $T_{rowmax}$, which is calculated as follows:

$$T_{rowmax} = \max\{T_{rowi}/\prod_{j<i} G_j\} \qquad (3)$$

where $G_j$ is the stride of convolution layer or pooling layer i. After flowing through these layer, H of feature map reduces so that latter layers have multiple times of cycle to process $K$ rows. The throughput, which is defined as frames per second, is calculated as follows:

$$throughput = f/(H_0 \times T_{rowmax}) \qquad (4)$$

where $H_0$ is the number of input activation rows and $f$ is the frequency of the DSP. To conclude, in order to get optimal throughput, the computation time of each layer should be balanced, which can be adjusted by the channel parallelism $C'$ and $M'$ for each convolution layer engine.

## 3.3. Convolution Layer Engine

A convolution layer engine consists of a processing element array, a weight buffer, an activation buffer and a controller shown in Fig.3 and ROMs for storing bias and shift bits.

Fig.3(a) shows the design of a processing element array with $M' \times C' \times R \times S$ multipliers. $C' \times R$ of input activations feed into multipliers in each cycle and weights keep their values until computation of current channels finishes. Multiplication results of different input channel are added together first and produced partial sums of the same kernel row are accumulated in a pipeline way. Such dataflow is called weight stationary, which reuses weights for at least one activation row and reuses activations for a kernel. Partial sums of all kernel rows are added by adder trees and stored in psumSpad temporarily, waiting for accumulation with partial sums of other groups of input channels. In order to improve weight data reuse further, $K$ rows of partial sums are computed before loading new weights, so that size of activation buffer and psumSpad increases. Signals such as zeroMac, flush and rowSel are generated by the controller, for handling zero padding of four directions automatically.

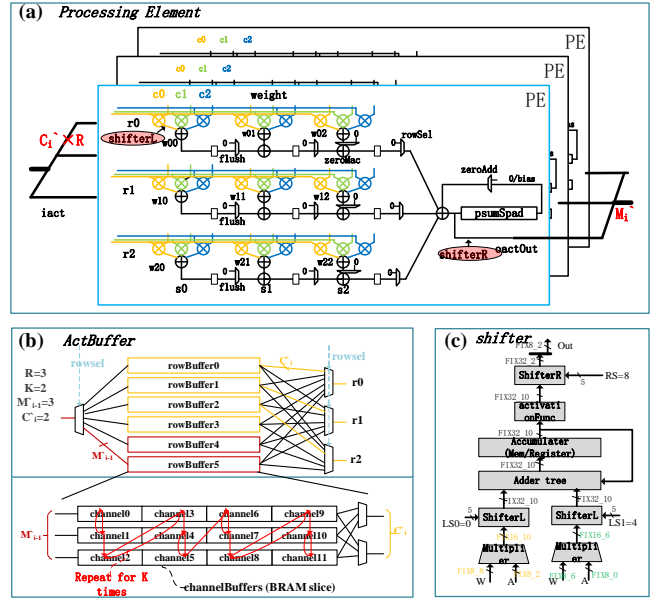

Fig.3 Convolution layer engine design

Activation buffer is designed for buffering output activations for previous layer and providing input activations for current layer. To keep processing elements busy, input and output widths of the activation buffer should correspond to the input parallelism $C'_i \times R_i$ for the current layer and the output parallelism $M'_{i-1}$ for the previous layer. We elaborate the activation buffer design to support the case when $C'_i$ and $M'_{i-1}$ are quite different. Considering stride is 1, to support simultaneous writing and reading, the activation buffer has $R + 2K - 1$ rowBuffers, of which $R + K - 1$ rows for reading and $K$ rows for writing. Each rowBuffer has $\max(C'_i, M'_{i-1})$ channelBuffers. Reading sequence for these channelBuffers is complicated and carefully processed by the appropriate address generator.

Weight and activation are computed as fix-points in this design. Quantization from floats to fixed-points naturally introduces precision decrease, but this can be alleviated by applying different fixed-point formats across layers and channels. Our architecture supports such channel-wise different fixed-point formats. As Fig.3(c) shows,

**Algorithm 1 Allocate computation resources**

1: compute the number of MAC computation $\pi_i$ for each layer:
$\pi_i = H_i W_i R_i S_i C_i M_i$
2: compute the multipliers required by each layer for best performance: $\hat{\theta}_i = \pi_i \times \Theta / \sum \pi_i$
3: assign multipliers to layer: $\theta_i = \lfloor \hat{\theta}_i / R_i S_i \rfloor \times R_i S_i$
4: while $\sum \theta_i \leq \Theta$:
5:     select layer j with maximum $\pi_i / \theta_i$
6:     if $\sum \theta_i + R_i S_i \leq \Theta$:
7:         $\theta_i = \theta_i + R_i S_i$
8:     else break
9: decompose $\theta_i$ into $C'_i$ and $M'_i$

**Algorithm 2 Allocate BRAMs considering bandwidth**

1: Initialize $K_i = 1$ for all layers
2: Compute weight access for each layer:
$\omega_i = H_i R_i S_i C_i M_i / K_i$
3: Compute the bandwidth required by all layers:
$B = (f / H0 \times Trowmax) \times \omega i$
4: While $B > \beta$:
5:    Assign $a_i$ BRAM for activation buffer of layer i:
$a_i = K_{i-1} + R_i + G_i(K_i - 1)$
6:    Compute total BRAM: $A = \sum a_i$
7:    choose layer i with max $\omega_i$
8:    if $A + G_i < \alpha$:
9:      $K_i = K_i + 1$
10:   else: break
11:   update $\omega_i$ and B

multiplication results of different fixed-point formats are aligned by left shifters before added together. To avoid overflow, partial sums are 32-bit for 8-bit activation and weight. While transferring to output activations, partial sums should be right shifted and truncated for scaling down.

## 4. RESOURCE ALLOCATION FRAMEWORK

The proposed framework can customize flexible pipeline accelerator according to different hyperparameters of CNN models and available hardware resources on FPGA. The parameterized convolution layer engine is developed in RTL, so that we can adjust their hardware resources just by a few parameters. The top module consists of instantiation of all layers and other components such as DDR interface.

### 4.1. Computation Resource

The multipliers are implemented only using DSPs, and one DSP48E1 slice of Xilinx FPGA has input bitwidth of 25-bit and 18-bit, which supports 1 multiplication for 16-bit quantization or 2 multiplications for 8-bit quantization per cycle. The computation resource allocation assigns DSPs to each convolution layer engine by deciding the parallelism $C'$ and $M'$, according to their computation workload. In order to achieve higher performance, computation cycles of all layers should be well balanced and DSPs should be utilized as many as possible.

Algorithm 1 summaries the allocation process of computation resources for a specific NN model. DSPs are pre-allocated based on the workload. If still existing available DSPs, assign more to slowest layer, iteratively.

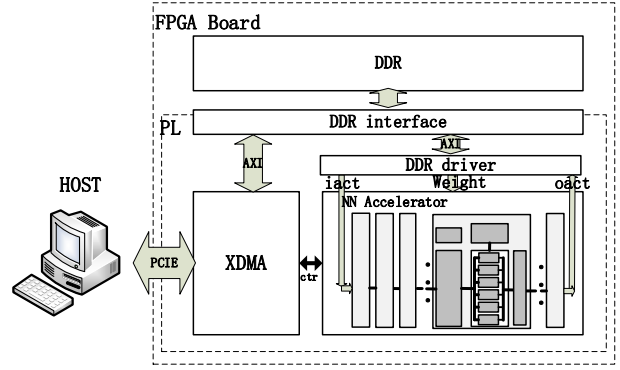

Fig.4 System implementation

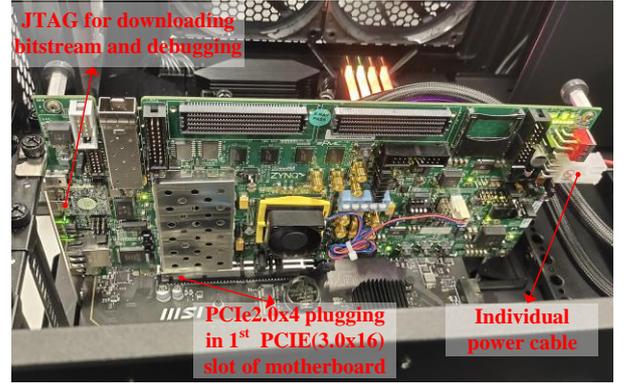

Fig.5 Hardware for experiment

### 4.2. BRAM and Off-chip Memory Bandwidth

Most of the DDR bandwidth in our pipeline accelerator is occupied by repeated loading of weights. We can increase row parallelism $K$ to improve weights reuse for mitigating the case when the DDR bandwidth is not sufficient. However, BRAM resources also need to be considered because more BRAMs are required to buffer input activation for the layer to be optimized. Algorithm II summarized the BRAM allocation. $\alpha$ and $\beta$ mean the total BRAM and DDR bandwidth of FPGA board. For layer i, activation buffer stores $R_i + K_{i-1} \times (K_i - 1) \times G_i$ rows.

## 5. RESULTS

### 5.1. Demo System

As Fig.4 shows, our FPGA based NN accelerator works with a host PC. Data and control signals are transferred through PCIE interface between host PC and the FPGA board. Activations and weights are buffered by the DDR and fetched on chip through the DDR interface IP by our developed DDR driver. After sending all weights of the NN model and input activations of several frames to DDR, host PC starts up the NN accelerator. A counter records the amount of output activations generated by the NN accelerator, which is checked by host PC. Host PC fetches output activations from FPGA boards and sends more input frames continuously. Fig.5 shows hardware for experiment.

Table I Utilization and performance comparison

| Model | VGG16 | | | | AlexNet | | ZF | | YOLO | |
|---|---|---|---|---|---|---|---|---|---|---|
| Reference | [1] | [2] | [3] | This Work | [3] | This Work | [3] | This Work | [3] | This Work |
| Frequency(MHz) | 150 | 100 | 200 | 200 | 200 | 200 | 200 | 200 | 200 | 200 |
| DSP (900) | 780 | 824 | 680 | 900 | 808 | 864 | 824 | 892 | 680 | 892 |
| LUT (218600) | 83% | 71% | 52% | 54% | 39% | 51% | 40% | 52% | 39% | 52% |
| FF (437200) | 29% | 28% | 14% | 34% | 12% | 36% | 12% | 35% | 11% | 44% |
| BRAM(545) | 89% | 83% | 99% | 74% | 56% | 84% | 61% | 58% | 61% | 76% |
| DSP Efficiency | 58.50% | 69.60% | 96.20% | 98.00% | 76.30% | 90.40% | 79.70% | 90.80% | 96.20% | 98.40% |
| Complexity(GOP) | 30.94 | 30.94 | 30.94 | 30.94 | 1.45 | 1.45 | 2.34 | 2.34 | 40.14 | 40.14 |
| Performance(GOPS,16b) | 137 | 230* | 262 | 353 | 247 | 312 | 263 | 324 | 234 | 351 |
| Performance(FPS,16b) | 4.4 | 7.4* | 8.5 | 11.3 | 170 | 230 | 112.2 | 138.4 | 5.8 | 8.8 |
| Performance(GOPS,8b) | 274 | / | 524 | 706 | 494 | 624 | 526 | 648 | 468 | 702 |
| Performance(FPS,8b) | 8.9 | / | 16.9 | 22.6 | 340 | 459 | 224.4 | 276.8 | 11.7 | 17.5 |
| Power(W) | 9.63 | 9.4 | 7.2 | 7.2* | 7.2 | 6.9* | / | 7.1 | / | 7.3 |
| Power Efficiency(GOPS/W,16b) | 14.2 | 24.4 | 36.4 | 49 | 34.3 | 43.3 | / | 45.6 | / | 48.1 |

## 5.2. Implementation Results

We have implemented AlexNet, VGG16, ZF and YOLO on Xilinx ZC706 board, which possesses 900 DSPs.

Table I summarize the results and comparison with the reference works. For VGG16, compared with recurrent architecture [1], the proposed work can achieve 1.6x higher DSP efficiency and utilize 120 more DSPs. Also working at higher frequency, the performance is 2.58x higher than [1], which indicates pipeline architecture has better optimization on DSP usage. Compared with fusion-based pipeline architecture [2], the proposed design with row-based pipeline architecture also has better DSP utilization and efficiency, so 1.53x higher performance, because the dataflow is easier for resources allocation. Besides, [2] adopts Winograd algorithm for convolution computing which reduce number of multiplications into one quarter, so the proposed work is 6.14x faster in terms of hardware performance. Compared with [3], the proposed work achieves much higher DSP utilization and efficiency for four NN models. Therefore, compared with [3] the performance for VGG16, AlexNet, ZF and YOLO are 35%, 26%, 23% and 50% higher on ZC706. This is the benefit coming from the flexible activation buffer design and more flexible resources allocation algorithm. The proposed design achieves better performance for all NN models, which proves the high flexibility. Notice that different from power meter result of [3], power consumption of our work is estimated by Vivado tool.

## 6. CONCLUSION

This work targets on high-throughput and flexible CNN computations based on FPGA. The layer-wise pipeline architecture is adopted, which is able to achieve high hardware resources utilization and efficiency on FPGA. Pipeline architecture and detailed convolution layer engine design are elaborated with high parameterization. Based on this, the framework can generate optimal design according to the features of various CNN model and FPGA devices. Benefited from flexible activation buffer, flexible resources allocation algorithm can achieve much higher DSP utilization. Besides, to support DSP utilization, dataflow is optimized to make use of BRAM to lower bandwidth requirement. Implemented on Zynq XC7Z045 the proposed design achieves much higher performance than the reference works on all four benchmarks, which proves the flexibility of the framework and the effectiveness of the above architecture optimization.

### Acknowledgements

This work was supported in part by JST, PRESTO Grant Number JPMJPR19M5, Japan.